\documentclass[12pt]{article}

\usepackage{amsfonts,amsmath,amsxtra}

\newcommand{\beq}[1]{\begin{equation}\label{#1}}
\newcommand\eeq {\end{equation}}
\newcommand\bqa {\begin{eqnarray}}
\newcommand\eqa {\end{eqnarray}}
\newcommand\pr {\partial}

\newcommand{\eq}[1]{eq.(\ref{#1})\ }
\newcommand{\bear}{\begin{array}}
\newcommand{\enar}{\end{array}}

\begin{document}

\def\t{\theta}
\def\T{\Theta}
\def\w{\omega}
\def\ov{\overline}
\def\a{\alpha}
\def\b{\beta}
\def\g{\gamma}
\def\s{\sigma}
\def\l{\lambda}
\def\wt{\widetilde}


\hfill{ITEP-TH-21/04}

\vspace{5mm}

\centerline{\bf \Large Expansion in Feynman Graphs as Simplicial
String Theory}

\vspace{3mm}

\centerline{{\bf Emil T.Akhmedov
}\footnote{email:{akhmedov@itep.ru}}}

\centerline{117259, Moscow, B.Cheremushkinskaya, 25, ITEP, Russia}

\centerline{and}

\centerline{141980, Moscow Region, Dubna Institute for Advanced
Study, JINR, Russia} \vspace{3mm}

\begin{abstract}
We show that the series expansion of quantum field theory in the
Feynman diagrams can be explicitly mapped on the partition
function of the simplicial string theory --- the theory describing
embeddings of the two--dimensional simplicial complexes into the
space--time of the field theory. The summation over
two--dimensional geometries in this theory is obtained from the
summation over the Feynman diagrams and the integration over the
Schwinger parameters of the propagators. We discuss the meaning of
the obtained relation and derive the one--dimensional analog of
the simplicial theory on the example of the free relativistic
particle.
\end{abstract}

\vspace{3mm}

{\bf 1.} There is a hope that the large $N$ Yang--Mills theory is
exactly equivalent to a string theory \cite{Polyakov:ez}. Such a
string theory if present can reveal the integrability of the large
$N$ Yang--Mills theory. Hence, the theory will help in explaining
the confinement phenomenon.

Despite the recent progress \cite{Maldacena:1997re,Akhmedov:un}
for the case of (super)conformal theories we still do not
understand the relation between gauge and string theories. We do
not understand which features of the relation are generic (persist
at least for the non--conformal and/or non--supersymmetric cases)
and which are specific for the concrete relation of
\cite{Maldacena:1997re}. This is due to the fact that there is no
explicit proof of the AdS/CFT correspondence.

In an attempt to understand the relation between field and string
theories in general we present, in this note, the explicit map
from the functional integral of the matrix field theory (at finite
$N$) onto the partition function of the simplicial string theory
--- the theory describing embeddings of the two--dimensional
simplicial complexes into the space--time of the field theory. Our
considerations are quite generic and can be applied to the
Yang--Mills theory. However, we consider the model example of the
matrix $\Phi^3$ theory whose interpretation on the string theory
side we understand best of all.

The map in question is given by a duality transformation. To some
extent this duality is the lattice analog of the T--duality map,
although we do not have any compact dimensions. Via this
transformation we map the summation over the Feynman diagrams and
the integration over the Schwinger parameters onto the sum over
the triangulations of the two--dimensional surfaces and the
integration over the invariant two--dimensional distances between
the vertices of the simplecial comlexes. This seems to be a
summation over {\it all} two--dimensional geometries and {\it all}
embeddings of the simplicial complexes into the space--time. To
understand this point we consider the toy example of the free
relativistic particle, for which we present a similar expression.
There, the summation over {\it all} one--dimensional geometries is
given by the summation over one--dimensional ``triangulations''
and integrations over the lengths between the vertices of the
``triangulations''. The integration over all positions of the
vertices gives the sum over {\it all} possible embeddings. The
resulting ``triangulated'' expression is {\it exactly equivalent}
to the relativistic particle path integral: No continuum limit
should be taken.

However, the complete understanding of the simplicial string
theory --- at least its possible continuum formulation, or may be
a continuum limit of it --- is still lacking. In particular, it is
possible that in the continuum formulation the theory describes
strings in the curved ${\rm AdS}_5$ space rather than in ${\rm
R}^4$ \cite{Gopakumar:2003ns}.

Anyway, as usual, the relation between two theories can be useful
to both of them. In fact, the map in question at least can give an
unambiguous way of formulating the simplicial string theory.
Particularly, the measure of integration and the two--dimensional
gravity action unambiguously follow from the matrix field theory.

The structure of the paper is as follows. In the second section we
present the map between the two theories. In the third section we
present interpretation of the resulting dual expression obtained
in the second section. In the fourth section we consider the
example of the free relativistic particle and present a simplicial
path integral for it. We conclude with the discussion in the fifth
section. In the Appendix we present a simple proof of the well
known combinatoric formulae \cite{Comb} for the Feynman integrals.
These formulae acquire a new meaning after the relation of the
field theory to the simplicial string theory is established.

{\bf 2.} Consider the matrix scalar field theory in the
$D$--dimensional Euclidian space:

\bqa Z = \int D \hat{\Phi}(x) \, \exp\left\{- \int d^D x \, N \,
{\rm Tr} \left[\frac12 \, \left|\vec{\pr} \hat{\Phi}\right|^2 +
\frac12 \, m^2 \, \left|\hat{\Phi}\right|^2 + \frac{\lambda}{3}
\hat{\Phi}^3\right] \right\}, \label{qft}\eqa where $\vec{\pr} =
(\pr/\pr x_1, \dots, \pr/\pr x_D )$, $\hat{\Phi}$ is $N\times N$
matrix field in the adjoint representation of $U(N)$ group:
$\Phi^{ij}$, $i,j = 1, \dots, N$. Note that we have re--scaled the
fields so that $\lambda$ is the 't Hooft coupling constant, but we
are {\it not} taking the large $N$ limit in this note.

The problems of this field theory, due to the sign indefiniteness
of the $\Phi^3$ potential, are not relevant for the most of our
further considerations: We consider the functional integral $Z$ as
a formal series expansion in the powers of $\lambda$. To deal with
connected graphs we consider $\log Z$.

It is well known that $\log Z$ can be represented as (see e.g.
\cite{BjorkenDrell} and \cite{t'Hooft}):

\bqa \log Z = \sum_{g=0}^{\infty} N^{\chi(g)} \sum_{V=0}^{\infty}
\lambda^V \, \sum_{\rm graphs; V,g \, fixed}
C_{\rm graph}(V,g) \times \nonumber \\
\times \left|\int_0^{+\infty} \prod_{n=1}^L d\alpha_n \, \int
\prod_{i=1}^V d^D \vec{y}_i \, \int \prod_{m=1}^L d^D \vec{p}_m \,
\exp\left\{- \sum_{l=1}^L\left[\frac{\alpha_l \, \left(\vec{p}_l^2
+ m^2\right)}{2} - {\rm i}\, \vec{p}_l \left(\Delta_l
\vec{y}\right) \right]\right\}\right|_{\rm graph}. \label{Feyn}
\eqa where $\vec{p}_l$ is the momentum running along the
propagator $l$; the propagators are written in the Schwinger
$\alpha$--representation; the first sum is taken over the genera
$g$ of the discretized closed two-dimensional surfaces represented
by the fat Feynman diagrams\footnote{Each member in the sum in
\eq{Feyn} is represented by the fat three--valent (three links are
entering each of the $V$ vertices) graph. Such a closed graph
represents a vacuum amplitude of the theory in \eq{qft}. The
generalization of our considerations for the correlators --- open
graphs --- is straightforward.} \cite{t'Hooft}; the second sum is
taken over the number $V$ of the insertions of
Tr$\hat{\Phi}^3(\vec{y}_i)$ vertices; $\chi(g) = V - L + F$ is the
Euler characteristic corresponding to the genus $g$ diagram in the
sum with $V$ vertices, $L$ propagators and $F$ faces\footnote{Do
not confuse this number with the number $G$ of the momentum loops
of the diagram; $F$ is the number of the closed index loops of the
{\it fat} Feynman diagram.}; the third sum is taken over graphs
(various Wick contractions) with fixed number $V$ of vertexes
Tr$\hat{\Phi}^3(\vec{y}_i)$ with fixed $g$; $\Delta_l \vec{y}$ is
the difference of the target space positions of the ends of the
$l$--th propagator; $C_{\rm graph}(V,g)$ are some combinatoric
constants.

For the general $D$ most of the integrals under the sum in
\eq{Feyn} are divergent. One of the types of the divergences is
proportional to the volume of the space--time and is just due to
the translational invariance. To get rid of this divergence we can
skip one of the $L$ integrations over the momenta. Another type of
the divergences are the standard UV divergences of quantum field
theory. We discuss them below.

 We are going to perform a transformation over
\eq{Feyn}. The same kind of transformation is performed in
\cite{Ivanenko:ya} and is referred to as duality on the lattice.
As well somewhat similar transformation is made in
\cite{Isaev:2003tk} and relates some types of the Feynman diagrams
of the $\Phi^3$ theory to the amplitudes in conformal Quantum
Mechanics.

To do this transformation let us perform the integration over the
$y$'s. Then each term under the sum and integration over
$\alpha$'s is represented as the {\it finite} function:

\bqa I(L,V, \{\alpha\}, {\rm graph}) = \int \prod_{m=1}^L d^D
\vec{p}_m \, \prod_{i=1}^V \delta\left(\sum^3_{l(\to i)}
\vec{p}_l\right) \, \exp\left\{- \sum_{l=1}^L\frac{\alpha_l \,
\left(\vec{p}_l^2 + m^2\right)}{2} \right\},\eqa where in each of
the $V$ $\delta$--functions the sum goes over the three links
terminating on each of the $V$ vertices. These are momentum
conservation conditions at each vertex. The UV divergences in the
diagrams appear after the integrations over the $\alpha$'s. To
perform the transformation in question we consider integrand
expressions, because we would like to show that this
transformation gives a non--trivial relation between the two
partition functions rather than a formal map from one infinite
number onto another. At the same time the divergences have a clear
physical meaning on the both sides of the realtion as is argued in
the next section.

The conditions which are imposed by the $V$ $\delta$--functions
are usually solved via $G = L - V + 1$ independent momenta running
along the loops of the diagram. However, we are going to solve
them via the dual graph to the Feynman diagram under
consideration. The dual graph consists of the vertices sitting in
the centers of the faces of the Feynman diagram and its links are
passing through the centers of the propagators of the Feynman
diagram. Thus, dual graph to a three--valent Feynman diagram
represents the triangulation of a two--dimensional surface: The
faces of the dual graphs are triangles.

Then each of the $L$ momenta $\vec{p}_l$ obeying conditions
$\sum_{l(\to i)}^3 \vec{p}_l = 0$ can be represented
as\footnote{Note that the momenta under the sums $\sum_{l(\to
i)}^3 \vec{p}_l = 0$ have alternating signs: Some momenta are
entering the vertex, while the others are exiting from it. This
obviously means that the links of the Feynman diagram have
orientations. Hence, the links of the dual graph should have
synchronized orientations (with the Feynman diagram) which will
define in \eq{pxrel} which of the edge $x$'s enters with ``+'' and
which with the ``-'' sign in the corresponding $p$ so that all the
momentum conservation conditions are fulfilled \cite{Valera}.}:

\bqa \vec{p}_l = \Delta_l \vec{x} + \sum^{2g}_{s=1} \vec{\mu}_s
\omega_l^{(s)}, \label{pxrel}\eqa where $\Delta_l \vec{x}$ is the
difference of the target space positions of the ends of the link
$l$ of the dual graph (which is intersecting the $l$--th
propagator of the Feynman diagram); $\vec{\mu}_s$ are arbitrary
parameters and $\omega_l^{(s)}$ are $2g$ closed (but not exact)
one--forms on the genus $g$ simplicial complex defined by the dual
graph. To explain these observations let us point out that the
condition $\sum_{l(\to i)}^3 \vec{p}_l = 0$ is equivalent to the
$D$ two--dimensional $d\vec{p} = 0$ conditions on the lattice
\cite{Ivanenko:ya}. The solutions of these conditions are $\vec{p}
= d\vec{x} + \sum^{2g}_{s=1} \vec{\mu}_s \omega^{(s)}$,
$d\omega^{(s)} = 0$ for all $s$ whose lattice expression is
\eq{pxrel}.

Using this solution we obtain:

\bqa\log Z = \sum_{g=0}^{\infty} N^{\chi(g)} \sum_{V=0}^{\infty}
\lambda^V \, \sum_{\rm graph; V,g fixed} C'_{\rm graph}(V,g) \,
\times \left|\int \, \prod_{s=1}^{2g} d^D\vec{\mu}_s \,
\left[\det\left(\sum_{n,m=1}^L \omega_n^{(s)}\,
\omega_m^{(s')}\right)\right]^{D/2} \times\right. \nonumber \\
\left. \times \int_0^{+\infty} \prod_{n=1}^L d\alpha_n \, \int
\prod_{a=1}^F d^D \vec{x}_a \, \exp\left\{- \sum_{l=1}^L
\frac{\alpha_l}{2} \, \left[\left(\Delta_l \vec{x}\right)^2 +
\left(\sum^{2g}_{s=1} \vec{\mu}_s \omega_l^{(s)}\right)^2 +
m^2\right]\right\} \right|_{\rm graph}
= \nonumber \\
= \sum_{g=0}^{\infty} N^{\chi(g)} \sum_{V=0}^{\infty} \lambda^V
\sum_{\rm graph; V,g fixed}\, C'_{\rm graph}(V,g) \times \nonumber
\\ \times \left| \int_0^{+\infty} \prod_{n=1}^L
\frac{d\alpha_n}{\alpha_n^{g \, D}} \, e^{-\frac{\alpha_n \,
m^2}{2}} \, \int \prod_{a=1}^F d^D \vec{x}_a \, \exp\left\{-
\sum_{l=1}^L \frac{\alpha_l}{2} \,\left(\Delta_l \vec{x}\right)^2
\right\}\right|_{\rm graph}, \label{simplex}\eqa where $C'_{\rm
graph}(V,g)$ are different from $C_{\rm graph}(V,g)$ by the $D/2$
power of the determinant of the matrix relating $p$'s and $x$'s in
\eq{pxrel}; $F$ is the number of the vertices (faces) of the dual
(Feynman) graph. In \eq{simplex} we have used the fact that
$\omega$'s are closed.

In the next section we interpret the expression in \eq{simplex} as
the simplicial string theory. In this context the summations over
the genera, triangulations and the integrations over the
$\alpha$'s give the summation over internal two-dimensional
geometries. The integration over the $x$'s --- positions of the
vertices --- gives the summation over the embeddings.

It is worth mentioning at this point that all our considerations
so far can be easily generalized to higher valent fat graphs (i.e.
to the matrix $\Phi^n$, $n \ge 4$ theory or to non--Abelian gauge
theories). However, the resulting dual graphs in these cases
contain more complicated simplexes than just triangles
\cite{Valera}.

{\bf 3.}  The definition of the simplicial string theory is well
known \cite{simp}. We present it here to make the interpretation
of \eq{simplex} obvious. First, the internal metric on a
simplicial complex is given by:

\bqa ||h_{\alpha\beta}||_\nabla = ||\vec{e}_\alpha \,
\vec{e}_\beta||_\nabla = \left(
\begin{array} {c c}
\vec{e}_1^2 & \vec{e}_1\, \vec{e}_2 \\ \vec{e}_1\, \vec{e}_2 &
\vec{e}_2^2
\end{array} \right) = \left(
\begin{array} {c c}
e_1^2 & \frac12\,\left[e_1^2 + e_2^2 - e_3^2\right] \\
\frac12\,\left[e_1^2 + e_2^2 - e_3^2\right] & e_2^2
\end{array} \right), \eqa where $\vec{e}_{\alpha}$, $\alpha = 1,2 $
are two--dimensional vectors which are setting the zweibein. They
are along two edges of each triangle $\nabla$ of the simplicial
complex. As well $e_{1,2,3}$ are lengths of the three edges of
these triangles. Second, the external metric on a simplicial
complex is given by:

\bqa ||G_{\alpha\beta}||_\nabla = \left(
\begin{array} {c c}
\left(\Delta_1 \vec{x}\right)^2 & \Delta_1 \vec{x}\, \Delta_2 \vec{x} \\
\Delta_1 \vec{x}\, \Delta_2 \vec{x} & \left(\Delta_2
\vec{x}\right)^2
\end{array} \right) = \nonumber \\ = \left(
\begin{array} {c c}
\left(\Delta_1 \vec{x}\right)^2 & \frac12\, \left[\left(\Delta_1
\vec{x}\right)^2 + \left(\Delta_2
\vec{x}\right)^2 - \left(\Delta_3 \vec{x}\right)^2\right] \\
\frac12\, \left[\left(\Delta_1 \vec{x}\right)^2 + \left(\Delta_2
\vec{x}\right)^2 - \left(\Delta_3 \vec{x}\right)^2\right] &
\left(\Delta_2 \vec{x}\right)^2
\end{array} \right). \eqa $\Delta_{1,2,3} \vec{x}$ are differences
of the target space positions of the vertices of each triangle of
the simplicial complex. Hence, the discretization of the string
theory action is as follows:

\bqa S = \int d^2\sigma \, \sqrt{h} \, h^{ab} \, \pr_a \vec{x} \,
\pr_b \vec{x} => \sum_{\nabla} \sqrt{h_\nabla} \, {\rm Tr}
\left(||h||_\nabla^{-1} \, ||G||_\nabla \right) = \nonumber \\
= \sum_{\nabla} \frac{\left[\left(\Delta_1 \vec{x}
\right)^{2\phantom{\frac12}} \left(e_2^2 + e_3^2 - e_1^2\right) +
\left(\Delta_2 \vec{x} \right)^2 \, \left(e_1^2 + e_3^2 -
e_2^2\right) + \left(\Delta_3 \vec{x} \right)^2 \, \left(e_1^2 +
e_2^2 -
e_3^2\right)\right]_{\nabla}}{2\,\sqrt{h_{\nabla}}}.\label{1}\eqa
Here the sum is going over all triangles of a simplicial complex
and

\bqa h_{\nabla} = \frac14\left[\left(e_2^2 + e_3^2 -
e_1^2\right)^{\phantom{\frac12}} \left(e_1^2 + e_3^2 -
e_2^2\right) + \left(e_2^2 + e_3^2 -
e_1^2\right)^{\phantom{\frac12}} \left(e_1^2 + e_2^2 -
e_3^2\right) + \right. \nonumber \\ \left. + \left(e_1^2 + e_2^2 -
e_3^2\right)^{\phantom{\frac12}} \left(e_1^2 + e_3^2 -
e_2^2\right) \right]_{\nabla} \label{2}\eqa is the determinant of
the internal metric. Thus, it is natural to define the partition
function of the simplicial string theory as:

\bqa Z_{sst} = \sum_{\rm Triangulations} \int \left[d
e\right]_{\rm Triangulation} \, e^{-S(e)} \, \left| \int
\prod_{a=1}^F d^D \vec{x}_a \, \exp\left\{- \sum_{l=1}^L
\frac{\alpha_l(e) \, \left(\Delta_l
\vec{x}\right)^2}{2}\right\}\right|_{\rm triangulation}, \label{3}
\eqa where $F$ is the number of vertices of the triangulation
under the sum; $\alpha_l(e)$, as follows from \eq{1} and \eq{2},
are the positive functions of the lengths of the edges of the two
triangles glued together via the link $l$. What is left to be
defined is the measure $\left[d e\right]$ and the weight $S(e)$ of
the summation over the two--dimensional geometries. If we would
like to integrate over the $e$'s themselves we have to impose the
triangle inequalities into the measure to keep the metric positive
defined.

  Now we can point out the equivalence between \eq{simplex} and
\eq{1}---\eq{3} with the suitable choice of $\left[d e\right]$ and
$S(e)$. In fact, the measure and the weight for the summation over
the two--dimensional geometries in \eq{simplex} unambiguously
follows from the matrix field theory. This measure is very natural
because the integration goes over the $\alpha$'s rather than $e$'s
which demand triangle inequalities to be imposed \cite{Valera}.
However, the expressions for the discretized versions of the
standard gravity actions in  terms of the $\alpha$'s are not
known. This explains the reason why usually in the formulation of
the simplicial string theory one is trying to express everything
through the $e$'s rather than the $\alpha$'s \cite{simp}.

 It is worth mentioning at this point that the UV divergences of
the quantum field theory in \eq{qft} acquire a clear
interpretation in the simplicial string theory description. These
divergences are just due to the boundaries in the space of all
metrics: I.e. due to the degenerate metrics, which correspond to
such situations when some of the triangles degenerate into links.
In this context it is interesting to understand the meaning of the
renormalization group within the simplicial string theory context
(see \cite{Akhmedov:1998vf} for the attempts of the explanation).

Note that \eq{3} and \eq{simplex} are explicitly reparametrization
invariant, because there the integration is going over all
reparametrization invariant two--dimensional lengths between the
vertices of the simplicial complexes and over the target space
positions $x$'s of the vertices rather than over the maps
\cite{Valera}. In the next section we will present similar
situation for the relativistic particle. After that we will be
ready for the discussion of the two--dimensional situation.

{\bf 4.} Consider the path integral for the relativistic particle:

\bqa G(\vec{x}, \, \vec{x}') = \int_{\vec{x}}^{\vec{x}'} D
\vec{x}(\tau) \, \int \frac{D e(\tau)}{VolDiff} \,
\exp\left\{-\frac12 \, \int_0^1 d\tau \,
\left[\frac{\dot{\vec{x}}^2}{e(\tau)} + m^2 \,
e(\tau)\right]\right\}\label{main1}\eqa with the measures
following from the norms:

\bqa||\delta \vec{x}(\tau)||^2 = \int_0^1 d\tau \, e(\tau)\,
[\delta \vec{x}(\tau)]^2 = T \int_0^1 d f [\delta \vec{x}(f)]^2
\quad {\rm and} \nonumber \\
||\delta e(\tau)||^2 = \int_0^1 d\tau \, e(\tau) \,
\left[\frac{\delta e(\tau)}{e(\tau)}\right]^2.\eqa The answer for
this path integral is \cite{Polyakov:ez}:

\bqa G(\vec{x}, \, \vec{x}') \propto \int_0^{\infty} \frac{d
T}{\sqrt{T}} \, \det^{\frac{1 - D}{2}}{\left(-\frac{1}{T^2} \,
\frac{d^2}{d f^2}\right)} \, \exp\left\{- \frac12 \,
\left[\frac{(\vec{x} - \vec{x}')^2}{T} + m^2 \, T\right]\right\}.
\label{sup}\eqa In the $\zeta$--function regularization we obtain:

\bqa G(\vec{x}, \, \vec{x}') = \int^{+\infty}_0 \frac{d
T}{T^{D/2}} \, \exp\left\{- \frac12\left[\frac{(\vec{x} -
\vec{x}')^2}{T} + m^2\, T\right]\right\} = \int \frac{d^D
\vec{p}}{(2 \, \pi)^D} \, \frac{e^{{\rm i}\, \vec{p}\, (\vec{x} -
\vec{x}')}}{\vec{p}^2 + m^2}.\label{solution}\eqa Thus,
$G(\vec{x}, \, \vec{x}')$ is the Green's function of the
Klein--Gordon equation.

At the same time there is another reparametrization invariant
regularization for the path integral of the relativistic particle:
The lattice regularization, where the lattice spacings are
reparametrization invariant one--lengths. In this regularization
naively one has ($T=\sum_{i=0}^M e_i$):

\bqa\det^{-\frac12} \left(-\frac{1}{T^2} \, \frac{d^2}{d
f^2}\right) = \int D \lambda(\tau) \, \exp\left\{-
\int_0^1\frac{\dot{\lambda}^2(\tau)}{2 \, e(\tau)} \,
d\tau\right\} => \nonumber
\\ =>
\int \, \sqrt{e_0}\, \prod_{i=1}^M \sqrt{e_i} \, d \lambda_i \,
\exp\left\{- \sum_{j=0}^M \frac{(\lambda_{j+1} - \lambda_j)^2}{2
\,
e_j}\right\} \propto \frac{\prod_{i=0}^M e_i}{T^{\frac12}}, \nonumber \\
{\rm where} \quad ||\delta \lambda||^2 = \int_0^1 d \tau \,
e(\tau) \, \left[\delta \lambda(\tau) \right]^2 => \sum_{0=1}^M
e_i \, \left(\delta \lambda_i \right)^2, \quad e_i = e(\tau_i)\,
\Delta\tau \eqa and $\delta\lambda_0 = \delta \lambda_{M+1} = 0$.
Note that $e$'s are invariant one--dimensional lengths. If this
expression for the determinant is substituted into \eq{sup} we
obtain:

\bqa G_{Latt}(\vec{x}, \, \vec{x}') = C_M \int \frac{d^D
\vec{p}}{(2 \, \pi)^D} \, \frac{e^{{\rm i}\, \vec{p}\, (\vec{x} -
\vec{x}')}}{\left(\vec{p}^2 + m^2\right)^{M - DM/2}}.\eqa Here
$C_M$ is some constant.

It seems that the naive lattice regularization does not work.
However, for this case we can present a multiple integral
expression which solves the Klein--Gordon equation \cite{Valera}.
To obtain it note that in contrast with respect to the evolution
type equations\footnote{Which obey:

\bqa K(\vec{x},\, \vec{x}'|(M+1)\,\Delta\tau) = \int \prod_{i=1}^M
d^D \vec{y}_i \, K(\vec{x},\, \vec{y}_1|\Delta\tau) \,
K(\vec{y}_1, \, \vec{y}_2|\Delta\tau) ... \, K(\vec{y}_M, \,
\vec{x}'|\Delta\tau).\eqa} the Green's function of the
Klein--Gordon equation has the following feature:

\bqa \int \prod_{i=1}^M d^D \vec{y}_i \, G(\vec{x},\, \vec{y}_1)
\, G(\vec{y}_1, \, \vec{y}_2) ... \, G(\vec{y}_M, \, \vec{x}') =
\int \frac{d^D \vec{p}}{(2\pi)^D} \frac{e^{{\rm i} \, \vec{p} \,
(\vec{x} - \vec{x}')}}{(\vec{p}^2 + m^2)^{M+1}} \neq G(\vec{x}, \,
\vec{x}'). \eqa However, it is easy to correct this formula in
such a way that the equality will hold. For example, we can put
(for any $M$):

\bqa G(\vec{x}, \, \vec{x}') \propto \int \prod_{i=1}^M d^D
\vec{y}_i \, \prod_{j=0}^M \frac{d^D \vec{p}_j}{(2\pi)^D} \, d
\left(\sum_{k=0}^M e_k\right) \, \exp\left\{\sum_{m=0}^{M}
\left[{\rm i}\, \vec{p}_m\, \left(\vec{y}_{m+1} - \vec{y}_m\right)
- \frac12 \, \left(\vec{p}_m^2 + m^2 \right) \, e_m
\right]\right\} \propto \nonumber
\\ \propto \int
\prod_{i=1}^M d^D \vec{y}_i \, \frac{d \left(\sum_{j=0}^M
e_j\right)}{\prod_{n=0}^M e_n^{D/2}}\, \exp\left\{- \frac12 \,
\sum_{m=0}^{M} \left[\frac{\left(\vec{y}_{m+1} -
\vec{y}_m\right)^2}{e_m} + m^2 \, e_m \right]\right\}.
\label{pres}\eqa where $\vec{y}_0 = \vec{x}$, $\vec{y}_{M+1} =
\vec{x}'$. In this formula we take the integral over the moduli
$\sum_{i=0}^M e_i = T$ rather than over all $e$'s and the
expression under this integral depends on $T$ rather than all
$e$'s separately. The latter fact can be seen explicitly after the
integration over $y$'s.

The expression in \eq{pres} seems to be a good candidate for the
``proper discretization'' of the relativistic particle path
integral. However, due to the integration over $T$ rather than
each separate $e$'s this integral does not seem to have a good
local field theory interpretation. Note that obviously:

\bqa \int d \left(\sum_{i=0}^M e_i\right)\dots \neq \int
\frac{\prod_{i=0}^M d e_i}{VolDiff} \dots\eqa for any $M$. This is
the main difference from the above case of the relativistic
particle {\it path integral} when the limit $M=\infty$ is
appropriately taken and the $\zeta$--function regularization
instead of the lattice one is applied.

To obtain the integration over all $e$'s let us perform the
following trick. Consider the equality:

\bqa\frac{1}{\vec{p}^2 + m^2} \propto \sum_{L=0}^{\infty}
\frac{(-1)^{L}}{L!} \, \int_0^{+\infty} \prod_{l=1 }^{L} \frac{d
e_l}{e_l} \, \exp\left\{-\frac{(\vec{p}^2 + m^2)}{2} \,
\sum_{n=0}^L e_n\right\}.\eqa Then it is possible to write (in
each term under the sum $\vec{y}_0 = \vec{x}$, $\vec{y}_{L+1} =
\vec{x}'$):

\bqa G(\vec{x}, \, \vec{x}') = \sum_{L=0}^{\infty}
\frac{(-1)^{L}\, C_L}{L!} \, \int_0^{+\infty} \prod_{n=1}^{L}
\frac{d e_n}{e^{D/2+1}_n} \int \prod_{i=1}^L d^D \vec{y}_i \,
\exp\left\{- \frac12 \, \sum_{l=0}^{L} \left[\frac{\left(\Delta_l
\vec{y}\right)^2}{e_l} + m^2 \, e_l \right]\right\},
\label{main}\eqa where $C_L$ are constants. The enumeration of the
links in this one--dimensional case coincides with the enumeration
of the vertices: $\Delta_l \vec{y} = \vec{y}_{l+1} - \vec{y}_l$,
$l=i$ and in the one--dimensional (open path) case $L+1 = V$. If
we take the integrals over $y$'s instead of $p$'s then the
conditions are $\Delta_i \vec{p} = \vec{p}_{i+1} - \vec{p}_i = 0$
(for all $i$) whose continuum limit analogs are $d\vec{p} =
\pr_\tau \vec{p}\, d\tau = 0$. The solution of the latter on the
one--dimensional interval is $\vec{p}(t) = const$. That is the
reason why, unlike the two--dimensional case in \eq{simplex}, in
the one--dimensional situation we do not have a non--trivial
expression for the Green's function through the $\vec{x}_a$'s.

However, the formula in \eq{main} is in many respects very similar
to the two--dimensional expression in \eq{simplex}. In fact, it
contains the summation over all discretizations of the
world--trajectory (which are one--dimensional triangulations) and
the integration over all one--dimensional distances between the
vertices $y_i$'s (which is the integration over the $\alpha$'s).
In the one--dimensional case $\alpha(e) = e$. The summation over
the embeddings is given by the integration over all possible
positions of the vertices ($y$'s).

{\bf 5.} Thus, we find that the log of the functional integral for
the matrix quantum field theory can be represented as the
partition function of the first quantized simplicial string
theory. In the latter we sum over all possible embeddings of all
possible simplicial complexes into the target space. Instead of
the summation over the two--dimensional metrics we sum over all
possible triangulations and invariant two--dimensional distances
between the vertices of the simplicial complexes. Both of them
seem to be summations over all two--dimensional geometries. At the
same time the action describing embeddings of the simplicial
complexes appears to be the discretization of the standard
Polyakov action for the relativistic string theory in the flat
space \cite{Polyakov:ez}.

In an attempt to understand the resulting simplicial string theory
we consider the relativistic particle case. Here we have two {\it
equivalent} expressions: \eq{main1} and \eq{main}. One of them
includes integration over all smooth one-dimensional metrics,
while the other expression contains, in effect, integration over
all singular one--dimensional metrics. Both of them are containing
summations over all one--dimensional geometries with the fixed
topology (open paths). Note that it is not necessary to take a
continuum limit in \eq{main} to obtain the correct solution to the
Klein--Gordon equation. It is not even clear how to take a
continuum limit in the expression like \eq{main}. In fact, taking
$L=\infty$ does not mean the continuum limit.

 Similarly to the one--dimensional case the two--dimensional expression
in \eq{simplex} is explicitly reparametrization invariant and
seems to include the summation over all two-dimensional
geometries. Then it is tempting to find exactly equivalent to it a
continuum expression containing integration over all smooth
metrics. This looks like a crazy idea. At least there is no good
reason why $\lambda^V \, C_{\rm graph}(g,V)$ are the appropriate
constants for the equality to be true with a suitable measure for
the smooth metrics: There is no freedom for the choice of $C_L$'s
in \eq{main}.

Frankly speaking, we do not know whether the aforementioned
temptation is meaningful or that it is necessary to take a
continuum limit, whatever it means. Possible reason for taking a
continuum limit can be as follows \cite{Valera}. In the
one-dimensional case we have a singled out point as the boundary
of the world trajectory rather than a curve
--- continuous sequence of points. Hence, the equation
for the path integral following from the variation of the boundary
point is just a differential equation. At the same time the
generalization of this differential equation to the
two--dimensional case is loop equation on the boundary curve.
Apart from that there seems to be another risk: The gravity action
after the change from $\alpha$'s to $e$'s can appear to be
non--local. However, we do not think that this is the case. In
fact, the non--locality if present should be rather trivial
because the change from $\alpha$'s to $e$'s is local (depends on
adjacent triangles) and the measure in \eq{simplex} depends on
$\alpha$'s locally (it is the product over the triangles). In any
case this question demands a separate investigation.

Anyway, we believe that the choice among the two possibilities can
be made after the derivation/understanding the meaning of the loop
equation or its discretized version for the open string theory in
\eq{simplex}.

I would like to acknowledge valuable discussions with A.Rosly,
A.Morozov, A.Mironov, N.Amburg, A.Isaev, A.Gorsky, T.Pilling,
Yu.Makeenko and D.Malyshev. Especially I would like to thank
V.Dolotin for intensive collaboration and sharing his ideas. This
work was done under the partial support of grants RFBR
02--02--17260, INTAS 03--51--5460 and the Grant from the President
of Russian Federation MK--2097.2004.2.

\vspace{5mm}

\underline{\bf Appendix: Calculations of the graphs.} In this
Appendix we sketch a proof of some combinatorial formulae for the
Feynman diagrams which have a clear meaning within the context of
the simplicial string theory. These formulae were proved in
\cite{Comb} using the electric net analogy \cite{Bjorken}. Our
proof is purely combinatoric and less tedious.

Consider a Feynman diagram in any scalar quantum field theory
(with standard --- polynomial in fields --- interactions) which
has $V_E$ external vertices, $V_I$ internal vertices and $L$
propagators. The positions of the external vertices are $z_a$, $a
= 1,..., V_E$. All propagators are written in the Shwinger
$\alpha$--representation. The expression for this diagram
$I(\vec{z}_1, ..., \vec{z}_{V_E})$ gives a quantum field theory
amplitude. We would like to represent the integrand expression
under the integration over the $\alpha$'s explicitly in a
combinatoric form \cite{Comb}.

To calculate this diagram let us present the recurrent relation
between the graphs \cite{Valera}. Consider a complete graph (all
vertices of which are connected by links to each other) with $V$
vertices. Assign to this graph the following expression:

\bqa F_V\left(\{\vec{z}_a\}\left|^{\phantom{\frac12}}\right.
\{\alpha_{(ab)}\}\right) = \prod_{a\neq b}^V
\frac{1}{\alpha_{(ab)}^{D/2}} \, \exp\left\{-
\frac{\left(\Delta_{(ab)} \vec{z}\right)^2}{2\,
\alpha_{(ab)}}\right\}. \eqa Let us take the integral say over
$\vec{z}_V$. The result is ($\beta = 1/\alpha$):

\bqa \int d^D \vec{z}_V \,
F_V\left(\{\vec{z}_a\}\left|^{\phantom{\frac12}}\right.
\{\beta^{-1}_{(ab)}\}\right) = \frac{\left(\prod_{a \neq b}^{V}
\beta_{(ab)}\right)^{D/2}}{\left(\prod_{a' \neq b'}^{V-1}
\tilde{\beta}_{(a'b')}\right)^{D/2}} \,
\frac{1}{\left(\sum_{c'=1}^{V-1} \beta_{(c'V)} \right)^{D/2}} \,
F_{V-1}\left(\{\vec{z}_{a'}\}\left|^{\phantom{\frac12}}\right.
\{\tilde{\beta}^{-1}_{(a'b')}\}\right),\eqa  where:

\bqa \tilde{\beta}_{(a'b')} = \beta_{(a'b')} + \frac{\beta_{(a'V)}
\, \beta_{(Vb')}}{\sum_{c'=1}^{V-1} \beta_{(c'V)}}, \quad a',b' =
1,..., V-1.\eqa Along this way we can obtain expression for any
graph. In fact, choosing big enough $V$, taking $\beta \to 0$ (or
$\alpha \to \infty$) for the missing links in the graph and making
the appropriate number of the integrations over $z$'s, we can
always do that. The resulting expression has a clear combinatoric
representation \cite{Comb}. This expression is easy to see by
induction from the presented here formulae. In particular, the
resulting expression for the aforementioned Feynman diagram is
\cite{Comb}:

\bqa I\left(\vec{z}_1,... \vec{z}_{V_E}\right) \propto
\int_0^{+\infty} \prod_{l=1}^{L} d\beta_l \, \beta_l^{D/2-2} \,
\frac{1}{\Delta(\beta)^{D/2}} \, \exp\left\{- \sum_{n=1}^L
\frac{m^2}{2\, \beta_n} - \frac{P(\beta, \vec{z})}{4}
\right\}.\eqa Here:

\bqa \Delta(\beta) = \sum_{t_1} \prod^{V_I} \beta \nonumber
\\ P(\beta, \vec{z}) = \frac{\sum_{t_2} \left(\prod^{V_I+1}
\beta\right)
\left(\Delta_{t_2}\vec{z}\right)^2}{\Delta(\beta)},\eqa where in
the first expression the sum is going over all so called
dual--trees $t_1$ of the diagram, while in the second expression
the sum is going over all dual--2--trees $t_2$ of the diagram. In
these expressions we take products of $\beta$'s along the
corresponding dual--co--trees and dual--co--2--trees
correspondingly; $\Delta_{t_2}\vec{z}$ is the difference of the
positions of the two external vertices that come together in a
dual--2--tree $t_2$.

The definition of all these ``dual--(co)--(2)--trees'' is as
follows \cite{Comb}. The tree graph (not necessary connected)
obtained by shrinking $V_I$ lines of the diagram such that all
$V_I$ internal vertices merge with the external vertices, but that
no pair of external vertices become coincident, is called a
dual--tree; the set of $V_I$ shrunk lines a dual--co--tree. If we
shrink $V_I + 1$ lines so that not only all the internal vertices
merge with the external ones, but also exactly two external
vertices come together, then the resulting graph is a
dual--2--tree; the set of $V_I + 1$ shrunk lines a
dual--co--2--tree.

 The reason why we present these formulae here is the following.
The same kind of formulae can be written for the dual graph to a
Feynman diagram. For the dual graph such an amplitude has the
meaning of ether a scattering amplitude of closed strings or an
open string amplitude. Hence, for this dual graph $\Delta(\beta)$
and $P(\beta, \vec{z})$ are related to the determinant of the
discretized two-dimensional Laplacian (in curved metric) and
two-dimensional classical action (with the boundary conditions
given by $z$'s) correspondingly \cite{Valera}. But this is a theme
for a separate scientific investigation.


\begin{thebibliography}{99}

\bibitem{Polyakov:ez}
A.~M.~Polyakov,``Gauge Fields and Strings'', Harwood Academic
Publishers, (1987).


\bibitem{Maldacena:1997re}
J.~M.~Maldacena,
Adv.\ Theor.\ Math.\ Phys.\  {\bf 2} (1998) 231 [Int.\ J.\ Theor.\
Phys.\  {\bf 38} (1999) 1113] [arXiv:hep-th/9711200];
S.~S.~Gubser, I.~R.~Klebanov and A.~M.~Polyakov,
Phys.\ Lett.\ B {\bf 428} (1998) 105 [arXiv:hep-th/9802109];
E.~Witten, {\it Adv.Theor. Math. Phys.}, {\bf 2} (1998) 253.

\bibitem{Akhmedov:un}
E.~T.~Akhmedov,
Phys.\ Usp.\  {\bf 44} (2001) 955 [Usp.\ Fiz.\ Nauk {\bf 44}
(2001) 1005];
E.~T.~Akhmedov,
arXiv:hep-th/9911095.

\bibitem{Gopakumar:2003ns}
R.~Gopakumar,
arXiv:hep-th/0308184;
R.~Gopakumar,
arXiv:hep-th/0402063.

\bibitem{Comb} C.S.~Lam and J.P.~Lebrun, {\it Il Nuovo Cimento}, LIX A, 4, (1969)
397. See as well 
N.~N.~Bogolyubov and D.~V.~Shirkov, ``Quantum Fields''.

\bibitem{BjorkenDrell} J.D.~Bjorken and S.D.~Drell, ``Relativistic Quantum Fields'',
McGraw Hill, (1965).

\bibitem{t'Hooft} J.~'t Hooft, {\it Nucl. Phys}, {\bf B72} (1974) 461.



\bibitem{Ivanenko:ya}
T.~L.~Ivanenko and M.~I.~Polikarpov,
Nucl.\ Phys.\ Proc.\ Suppl.\  {\bf 26} (1992) 536;
M.~N.~Chernodub and M.~I.~Polikarpov,
arXiv:hep-th/9710205.

\bibitem{Isaev:2003tk}
A.~P.~Isaev,
Nucl.\ Phys.\ B {\bf 662} (2003) 461 [arXiv:hep-th/0303056].

\bibitem{Valera} I would like to thank V.V.~Dolotin for the
discussion on these points; E.T.~Akhmedov and V.V.~Dolotin to
appear.

\bibitem{simp} F.David, ``Simplicial Quantum Gravity and Random Lattices'',
hep--th/9303127; J.Ambjorn, M.Carfora and A.~Marzuoli, ``The
Geometry of Dynamical Triangulations'', hep-th/9612069;
H.W.~Hamber and R.M.~Williams, ``Gauge Invariance in Simplicial
Gravity'', hep--th/9607153.

\bibitem{Akhmedov:1998vf}
E.~T.~Akhmedov,
Phys.\ Lett.\ B {\bf 442} (1998) 152 [arXiv:hep-th/9806217];
E.~T.~Akhmedov,
arXiv:hep-th/0202055.

\bibitem{Bjorken} J.D.~Bjorken, Stanford Doctoral Thesis (1958).


\end{thebibliography}
\end{document}